\documentclass[doublecol,epsf]{epl2} 
\usepackage{bm}

\title{Deconfinement of Vortices with Continuously Variable Fractions of 
the Unit Flux Quanta in Two-Gap Superconductors}

\author{Jun Goryo\and Singo Soma\footnote{Present Affiliation; (C) HONDA Motor 
Co., Ltd.}\and Hiroshi Matsukawa}
\shortauthor{Jun Goryo, Singo Soma and Hiroshi Matsukawa}

\institute{                    
  \inst{1} Department of Physics and Mathematics, Aoyama Gakuin Univetsity - 5-10-1, Fuchinobe, Sagamihara, 229-8558, Japan\\
}

\abstract{
We propose a new stage of confiment-decofinment transition, which can be observed in laboratory.
In two-gap superconductors (SCs), two kinds of vortex exist 
and each of them carries a continuously variable 
fraction of the unit flux quanta $\Phi_0=h c / 2 e$. 
The confined state of these two is a usual vortex and stable in the low 
temperature region of the system under a certain magnetic field above $H_{c1}$. We see 
an analogy to quarks in a charged pion. An entropy gain causes two fractional vortices 
to be deconfined above a certain temperature. 
}

\begin{document}

\maketitle

\maketitle

\section{Introduction} Topological defects furnish fascinating problems 
in physics, and keep attracting a lot of concerns. 
In general, they lead the quantization of 
physical quantities. Vortex is one of the most important example of 
topological defects\cite{SC}. 
It is well known that in superconductors (SCs), the requirement 
that the order parameter is a single-valued function guarantees 
the quantization of the magnetic flux of a vortex in the unit of $\Phi_0=hc 
/ 2e$. 

The fractionalization of topologically quantized numbers have also been 
researched extensively in 
many different contexts \cite{fractionalization}. 
Vortices with fractional fluxes have been discussed in 
$p$-wave superfluid $^3$He \cite{half-flux-in-p} and $d$-wave   
SCs with a grain boundary \cite{Bailey-et-al} 
(See also a notice \cite{notice}). 
The fractionalization comes from unconventional structures of order 
parameters 
for these superfluid and SCs without any contradiction to their 
single-valued nature. 

Here we discuss vortex state in two-gap SCs \cite{two-gap} 
which has two superconducting gaps 
on two separate Fermi surfaces. 
The two-gap superconductivity can be seen in a lot of SCs in 
solid state and has been paid special attention. 
MgB$_2$ is a recent typical textbook\cite{MgB2}. 
The possibility of the two-gap 
superconducting state is also pointed out in 2H-NbSe$_2$\cite{NbSe2}, PrOs$_4$Sb$_{12}$\cite{PrOs4Sb12}, 
Y$_2$C$_3$\cite{Y2C3}, Ca-doped YBCO\cite{Ca-YBCO},  and liquid hydrogen \cite{hydrogen}. 
The fractionalization of the flux quanta has also been discussed in two-gap   
SCs\cite{Tanaka,Babaev}. Tanaka considered a wire ring and 
showed that a trapped 
flux in the ring can be an arbitrary fraction of $\Phi_0$ \cite{Tanaka}. 
Babaev considered a planer geometry and discussed the spontaneous fractional 
vortices associated with the Berezinskii-Kosterlitz-Thouless  (BKT) transition 
\cite{Kogut} in the weak coupling limit of an internal Josephson interaction 
with respect to the relative phase of the two gap functions\cite{Babaev}. 

Besides these pioneering works, searching for 
the appearance 
of the fractional vortices under other setups is challenging and important 
since it gives new viewpoints of the problem and also 
new possibility for experimental discovery. 
In this paper, we propose a novel mechanism for the appearance of fractional vortices in the two-gap 
SCs. We consider the system under an applied magnetic 
field and show that a vortex with the unit flux is divided into two vortices 
with continuously variable fraction of $\Phi_0$ by entropic effect. This 
phenomena can be seen as 
deconfinement of two fractional vortices. 
We see that in the confined state, two vortices behave like 
quarks in the charged pion \cite{Kogut}. 
Hence the present theory  gives rise to a new stage of confinement-decofinement 
transition, which can be examined in laboratory.

\section{Fractional vortices in a two-gap SC} Let us discuss the 
property of fractional vortices in a type-II two-gap SC. We consider two gap 
functions $\Delta_{\rm L}$ and $\Delta_{\rm S}$ ($|\Delta_{\rm L}| \geq |\Delta_{\rm S}|$) opened 
on two different Fermi surfaces. Their pairing symmetry does not matter. 
We can consider not only $s$-wave but 
also unconventional pairing states \cite{Sigrist-Ueda}. 
We apply a magnetic field along $z$-direction 
to a sample of thin film geometry, i.e., $d<min(\xi_z^{\rm L},\xi_z^{\rm S})$, 
where $d$ is the sample thickness and 
$\xi_z^{\rm L}$ and $\xi_z^{\rm S}$ are the coherence lengths of the gap functions along $z$-direction.  
The gap functions 
are then expressed as, 
$\Delta_{\rm L,S}({\bf r})=|\Delta_{\rm L,S}({\bf r})|\exp[-i \theta_{\rm L,S} 
({\bf r})], $
where $ {\bf r} $ is the two dimensional vector in the $xy$-plane. 
The Ginzburg-Landau (GL) free energy for two-gap system  
is straight extension of conventional one and given in Ref. \cite{Zhitomirski-Dao}. 
We consider the GL free energy in the London limit, namely, we 
neglect the spacial dependence of $|\Delta_{\rm L}({\bf r})|$ and 
$|\Delta_{\rm S}({\bf r})|$ and obtain  
\begin{eqnarray} 
F_{\rm GL}&=&d \int d^2 r \left[\frac{{\bf H}^2}{8 \pi} \right. 
+ \frac{1}{8\pi \lambda^2_{\rm L}(T)} \left({\bf A} - \frac{\Phi_0}{2 \pi} \bm\nabla 
\theta_{\rm L}\right)^2 
\nonumber\\ 
&&\left. + \frac{1}{8 \pi \lambda^2_{\rm S}(T)} \left({\bf A} - \frac{\Phi_0}{2 \pi} 
\bm\nabla \theta_{\rm S}\right)^2 \right.
\nonumber\\
&&
\left.-\Gamma(T) \cos(\theta_{\rm L} - \theta_{\rm S})  \right]. 
\label{London-limit}  
\end{eqnarray} 
The characteristic points of this free energy are the presence of two typical length scales, 
$\lambda_{\rm L,S}(T)=\left(4 e^2 K_{\rm L,S} |\Delta_{\rm L,S}(T)|^2/\hbar^2 
c^2\right)^{-1/2}$, 
where $K_{{\rm L,S}}$ are the coefficients of the gradient terms in GL free energy, 
and of the Josephson-type interaction with a temperature dependent coupling 
$\Gamma(T) \equiv 2 \gamma |\Delta_{\rm L}(T)||\Delta_{\rm S}(T)|$. 
This coupling causes the locking effect for the relative phase $\theta_{\rm 
L} - \theta_{\rm S}$. 
The equation of motion for the gauge field ${\bf A}$ is 
obtained from $\delta F_{\rm GL} / \delta {\bf A} =0$ as, 
\begin{equation} 
\lambda(T)^2 \bm\nabla \times {\bf H}=-{\bf A} + \frac{\Phi_{\rm L}(T)}{2 
\pi} \bm\nabla \theta_{\rm L}+ \frac{\Phi_{\rm S}(T)}{2 \pi} \bm\nabla 
\theta_{\rm S}, 
\label{eq-of-motion} 
\end{equation} 
where 
\begin{equation} 
\lambda^2(T) \equiv \left(\frac{1}{\lambda^2_{\rm 
L}(T)}+\frac{1}{\lambda^2_{\rm S}(T)}\right)^{-1} 
\label{lambda} 
\end{equation} 
is the London penetration depth, and 
\begin{eqnarray} 
\Phi_{\rm L,S}(T)&=&\Phi_0 \frac{\lambda^2(T)}{\lambda^2_{\rm L,S}(T)}, 
\label{fractional-flux}\\ 
\Phi_{\rm L}(T) &+& \Phi_{\rm S}(T)=\Phi_0. 
\label{combination} 
\end{eqnarray} 
The following boundary conditions are imposed,
\begin{eqnarray} 
\bm\nabla \times \bm\nabla \theta_{\rm L,S} ({\bf r})&=&2 \pi \hat{\bf z} 
\delta^2 ({\bf r} - {\bf R}_{\rm L,S}),  
\label{BC} 
\end{eqnarray} 
which would be relevant for the dilute vortex system with $H \simeq H_{c1}$.  
The singlevaluedness  of two gap functions is guaranteed 
except for ${\bf R}_{\rm L}$ 
and ${\bf R}_{\rm S}$ because of the $2\pi$ winding. 
From Eq. 
(\ref{eq-of-motion}), we obtain the London equation for the two-gap case 
\begin{eqnarray} 
&&\lambda^2(T) \bm\nabla \times \bm\nabla \times {\bf H}+{\bf H} 
\nonumber\\
&=&\Phi_{\rm L} (T)\hat{\bf z} \delta^2 ({\bf r} - {\bf R}_{\rm L}) + \Phi_{\rm S} (T) \hat{\bf z} 
\delta^2 ({\bf r} - {\bf R}_{\rm S}). 
\label{London-eq} 
\end{eqnarray} 
This equation tells us important implications. 
When ${\bf R}_{\rm L}={\bf R}_{\rm S}={\bf R}$, the equation reduces to the 
usual London equation $\lambda^2(T) \bm\nabla \times \bm\nabla \times {\bf 
H}+{\bf H}= \Phi_0 \hat{\bf z} 
\delta^2 ({\bf r} - {\bf R})$, and we have a vortex with the unit flux 
passing through 
${\bf R}$.  When ${\bf R}_{\rm L} \neq {\bf R}_{\rm S}$, however, we have 
two vortices with $\Phi_{\rm L}(T)$ through ${\bf R}_{\rm L}$ and 
$\Phi_{\rm S}(T)$ through ${\bf R}_{\rm S}$. 
It should be emphasized that $\Phi_{\rm L}(T)$ and $\Phi_{\rm S}(T)$ are fractions 
of $\Phi_0$ and its ratio 
is determined by the ratio $\lambda_{\rm L} (T)/ \lambda_{\rm S} (T)$ (See, Eqs. 
(\ref{lambda}), (\ref{fractional-flux}), and (\ref{combination})).  
When $\lambda_{\rm L}(T)/\lambda_{\rm S}(T)$ is limited to unity (such a limitation 
could come from the crystal symmetry, if we consider unconventional states 
in two dimensional 
irreducible representations in a system with single Fermi 
surface\cite{Sigrist-Ueda}), 
we have only the half flux $\Phi_{\rm L}(T)=\Phi_{\rm S}(T)=\Phi_0 / 2$, but there 
is {\it no} such a limitation 
in the present system and fractional vortices besides the half flux 
vortex are possible to exist 
when ${\lambda}_{\rm L}(T) \neq {\lambda}_{\rm S}(T)$.  It should be 
also emphasized that these fractional fluxes vary continuously 
with 
$|\Delta_{\rm L}(T)|/|\Delta_{\rm S}(T)|$, which determines the ratio 
$\lambda_{\rm L} (T)/ \lambda_{\rm S}(T)$.

There is a possibility to observe these fractional vortices 
when their separation 
is large enough.    
It would be small in the low temperarture region  
due to the Josephson type interaction in the GL free energy (\ref{London-limit}), 
but is possible to be large in the high temperature regime in order to 
gain entropy.  To check this scenario, we calculate the energy cost and 
the entropy gain for a certain configuration of two vortices with $\Phi_{\rm L}(T)$ and $\Phi_{\rm S}(T)$.

\section{Configuration energy} We can eliminate the gauge field 
from Eq. (\ref{London-limit}) by using the equation of motion 
(\ref{eq-of-motion}), and obtain the energy cost,
\begin{eqnarray} 
H_{\rm v}&=&d\int d^2 r \left[\frac{{\bf H}^2}{8 \pi} + \frac{\lambda^2(T)}{8 \pi} (\bm\nabla \times {\bf 
H})^2 
\right. 
\label{energy-cost1} \\ 
&+&\frac{K(T)}{2}\left\{ \bm\nabla (\theta_{\rm L} 
- \theta_{\rm S})\right\}^2 
\left. - \Gamma(T) \cos(\theta_{\rm L} - \theta_{\rm S}) \right], 
\nonumber
\end{eqnarray} 
where $K(T)\equiv \left\{\Phi_{\rm L}(T) \Phi_{\rm S}(T)\right\}/16 \pi^3 \lambda(T)^2$.
The vortex solution in the thin film geometry for the single gap 
SC has been investigated 
\cite{Peal}. The extension to the two-gap case is straightforward.  We 
introduce the effective London penetration depth in the thin film $\lambda_{\rm eff} (T)
=\lambda^2(T) / d$ \cite{Peal}.   
When $\lambda_{\rm eff}(T)$ is 
comparable to the system size,   
we obtain 
\begin{eqnarray} 
H_{\rm v} 
&\simeq&\sum_{i={\rm L,S}} \epsilon_i(T)-2 \pi K(T) \ln \frac{|{\bf 
R}_{\rm L} - {\bf R}_{\rm S}|}{\lambda_{\rm eff}(T)} +E_{\rm rel} ,  
\label{Hv}
\end{eqnarray} 
where 
$$
\epsilon_{\rm L,S}(T) \equiv \left\{\Phi_{\rm L,S}^2(T)/16 \pi^2 \lambda_{\rm eff}(T)\right\} \ln 
(\lambda_{\rm eff}(T)/\xi_{\rm L,S}(T))
$$ 
are the creation energy for L- and S-vortices, 
$\xi_{\rm L,S}(T) \equiv \hbar v_{\rm F}^{\rm L,S}/|\Delta_{\rm L,S}(T)|$ the coherence lengths of L- and 
S-gap functions, each of which should be in the same order of each vortex core, and $v_{\rm F}^{\rm L,S}$ the Fermi velocities.    
The logarithmic term in the {\it r.h.s} is the magnetic interaction between 
two vortices, which is repulsive and shows the logarithmic  behavior. The 
last term has a sine-Gordon form 
\begin{eqnarray} 
E_{\rm rel}&=&d \int d^2 r \left[\frac{K(T)}{2}\left\{ \bm\nabla (\theta_{\rm L} 
- \theta_{\rm S})\right\}^2\right.
\nonumber\\ 
&&\left.- \Gamma(T) \cos(\theta_{\rm L} - \theta_{\rm S}) \right]. 
\label{sine-Gordon} 
\end{eqnarray} 
When $\gamma=0$, $\Gamma(T)=0$ and $E_{\rm rel} \propto \ln |{\bf R}_{\rm L} - 
{\bf R}_{\rm S}|$ \cite{Babaev}. Such a case can be applied 
to the superconductivity in liquid hydrogen \cite{hydrogen}. 
We go beyond the weak coupling limit of $\gamma$, which 
would be significant for the two-gap superconductivities
in solid state, e.g., MgB$_2$ \cite{MgB2}. 

In Eq. (\ref{sine-Gordon}), the spatial dependence of the relative phase 
looses the elastic energy. 
The Josephson-type term favors to be $\theta_{\rm L} - \theta_{\rm S} = 2 
\pi n$ for $\gamma>0$ and $\theta_{\rm L} - \theta_{\rm S} = (2 n + 1) \pi $ 
for $\gamma<0$. But 
we have the boundary condition (\ref{BC}). 
There should be a line-like singularity between separated 
vortices associated with a $2\pi$ jump of the relative phase. This "string" 
is the two-dimensional extension of the kink in the sine-Gordon model in one 
dimension \cite{kink}.  Therefore, for large  $\gamma$, 
\begin{equation} 
E_{\rm rel} = \epsilon_{\rm st} (T) L_{\rm st} + const., 
\label{energy-cost2} 
\end{equation} 
where $\epsilon_{\rm st}(T) \equiv d \sqrt{K(T)|\Gamma(T)|}/\pi+ {\cal{O}} (1 / \gamma)$ is the string creation energy 
per a unit length and 
$L_{\rm st} \geq |{\bf R}_{\rm L} - {\bf R}_{\rm S}|$ is the length of 
the string. 
To minimize $H_{\rm v}$, $L_{\rm st} \rightarrow |{\bf R}_{\rm L} - {\bf 
R}_{\rm S}|$. This indicates 
that there is a linear confinement potential between L- and S-vortices. 
We have also the two-dimensional Coulomb type repulsion. 
Then, two fractional vortices is analogous to massive "quarks" with 
"fractional electric charge" $\Phi_{\rm L}$ and $\Phi_{\rm S}$, except 
for the spatial dimensionality \cite{Kogut}. 
The L-S confined state corresponds to the "charged pion" (total charge $+e$ 
of pion corresponds to 
the unit flux $\Phi_0=\Phi_{\rm L}+\Phi_{\rm S}$ of L-S confined state).

\begin{figure}
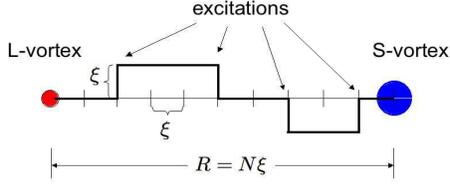

\onefigure[width=7cm,height=5cm]{st-ex.epsf}
\caption{A draw with four excitations of the string. 
The tick line denotes the string connecting 
two fractional vortices with a distance R.}
\label{fig:st-ex}
\end{figure}

\begin{figure}
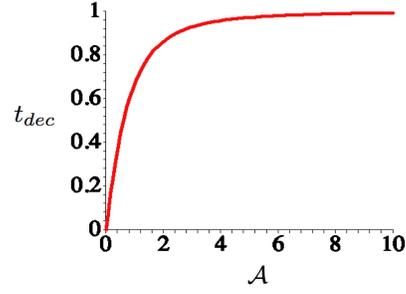

\onefigure[width=7.5cm,height=5cm]{t_dec.epsf}
\caption{${\cal{A}}$ dependence of $t_{dec}\equiv T_{dec}/T_c$, where $T_{dec}$ is the 
deconfinement temperature. }
\label{fig:t_dec}
\end{figure}

\begin{table}
\caption{Parameters which would be relevant for MgB$_2$ \cite{MgB2} }
\label{tab:table1}
\begin{center}
\begin{tabular}{cc||cc}
\hline
\hline
$\xi(0) / \lambda(0) $  & 0.32 &$\gamma$&1.6/eV/cell\\
$ d $  & 10 $\AA$ & $x$ &0.33\\
$\alpha$ & 2.2 &$ k $ & 2.1\\
\hline
\hline
\end{tabular}
\end{center}
\end{table}

\begin{figure}
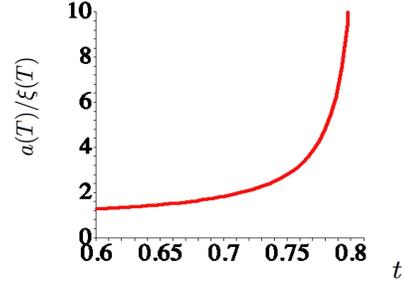

\onefigure[width=7.5cm,height=5cm]{a-plot.epsf}
\caption{$a(T)$ for ${\cal{A}}=1.6$. In this case, $t_{dec}\simeq 0.81$. Note that the unit of the vertical line is $\xi(T)=\xi(0)/
\sqrt{1-t}$. Here, we use the relation $a(0) \simeq 0.3\xi(0)$ obtained from Table \ref{tab:table1}.}
{\label{fig:a(T)} }
\end{figure}

\section{Entropy, confinement, and deconfinement}

Let us examine the thermodynamic stability of the separated state with a distance $R \equiv |{\bf R}_{\rm L} - {\bf R}_{\rm S}|=N \xi \neq 0$, where $\xi \equiv \max (\xi_{\rm L}, \xi_{\rm S})$ and we refer it as the cut-off length of this system at the short length scale. 
We introduce an artificial square lattice with a constant $\xi$ so that 
two vortices sit on one of the lattice axes. We discuss the free energy $F=H_{\rm v} - TS$, where $H_{\rm v}$ is given in Eqs. (\ref{Hv}) and (\ref{energy-cost2}). There are two contributions to the entropy $S=S_{\rm loc} + S_{\rm st}$, the former one comes from the location of two vortex cores $S_{\rm loc} \simeq 2 k_{\rm B} \ln (\Omega/\xi^2)$ ($\Omega$; the area of the system) and the latter the string configuration $S_{\rm st}$. 

The lowest energy configuration of the string is the straight line connecting two vortices, where $L_{\rm st}=R$. The kink-like configurations of the string would be thermally excited and each excitations contributes $\xi$ to $L_{\rm st}$ (See, Fig. \ref{fig:st-ex}). As shown in Fig. \ref{fig:st-ex}, there are two kinds of excitations one of which is directed to the "upward" and the other "downward". The numbers of the upward and downward excitations are equal each other so that the string is ended by two 
fixed points (vortices) and we denote them as $n$. 

We consider states with $L_{st} \stackrel{>}{\simeq} R$, i.e., $n \ll N$. 
By using the dilute approximation we obtain $S_{\rm st}=k_{\rm B} \ln {}_N C_n {}_{N-n} C_n \simeq k_{\rm B} (\delta R / \xi) \left[1 - \ln  \delta R/ 2 R \right]$, where $\delta R=2 n \xi$. We find the value of $\delta R$ which minimize $F$ at fixed $R$ is $\delta R_0(R)=2 R \exp\left[- \xi \epsilon_{\rm st} / k_{\rm B} T\right]$. Let us define a function $\tilde{F}(R) \equiv F(R,\delta R_0(R))$. 
We introduce $a(T)$ as the value of the vortex distance $R$ at the mimimum of $\tilde{F}(R)$ 
and it shows a diameter of the confined region of two fractional vortices. We obtain 
\begin{equation}
a(T)=\frac{2 \pi K(T)/\epsilon_{\rm st}(T)}{1 -  \frac{2 k_{\rm B} T}{\xi(T) \epsilon_{\rm st}(T)} \exp\left[-\frac{\xi(T) \epsilon_{\rm st} (T)}{k_{\rm B} T}\right]}. 
\label{a(T)}
\end{equation}
Remind that the solution of $1 - 2 /x \cdot \exp[-x]=0$ is $x={\rm LambertW}(2)\simeq 0.85$. Then, $a(T) \rightarrow \infty$ \cite{note-a} when 
\begin{equation}
k_{\rm B}T-\frac{\xi(T) \epsilon_{\rm st}(T)}{{\rm LambertW}(2)}=0.    
\label{T_dec}
\end{equation}
The temperature that satisfies Eq. (\ref{T_dec}) would correspond to the deconfinement temperature and we refer it as $T_{dec}$. We emphasize that {\it it is surely to be} $T_{dec}<T_c$, where $T_c$ is the gap opening temperature. The reason is that $\xi(T) \epsilon_{\rm st}(T)$ monotonically decreases to increase $T$ and vanishes at $T_c$. Then, the $l.h.s.$ of Eq. (\ref{T_dec}) changes its sign once between the zero temperature and $T_c$. This statement is 
crucial since vortices cannot exist above $T_c$ and the argument becomes inconsistent 
when $T_{dec}>T_c$. 

We can also show the presence of $T_{dec}$ when we consider states with $L_{\rm st}>>R$. 
Let us remind that the number of the configuration $N_{\rm con}=\mu^n$ for the $n$-step self-avoiding random walk, where $\mu$ is a constant and equal to 3 for the 2D square lattice case. 
For the string with sufficiently large $L_{\rm st}$, $N_{\rm con}$ would be well approximated by the random walk result and we obtain $N_{\rm con}\simeq\mu^{L_{\rm st}/\xi}$. Then, the entropy of the string is proportional to $L_{\rm st}$ \cite{Kogut} and the string free energy $F_{\rm st}\simeq \tau_{\rm st} (T) L_{\rm st}$, where the coefficient $\tau_{\rm st}(T)\equiv \left\{\epsilon_{\rm st}(T) - k_{\rm B} T/\xi(T)\cdot \ln \mu \right\} $ denotes the string tension and is positive in the low temperature region, i.e., fractional vortices are confined. Above a certain tempetarure, $\tau_{\rm st}(T)<0$, namely, the string looses its tension and deconfinement occurs. $T_{dec}$ is obtained from the equation 
$\tau_{\rm st}(T_{dec})=0$, which is equivalent to Eq. (\ref{T_dec}) when we take $\mu=\exp\left[{\rm LambertW}(2)\right]\simeq 2.35$.  

To obtain $T_{dec}$, we should know the temperature dependence of two gaps. By using GL equation for two-gap superconductor with sufficiently large $\gamma$ \cite{Zhitomirski-Dao}, we check that 
$|\Delta_{\rm L,S}(T)|/|\Delta_{\rm L,S}(0)|= k \sqrt{1-t}$, where $t\equiv T/T_c$ and $k$ is a constant which can be obtained numerically. Then,  Eq. (\ref{T_dec}) becomes ${\cal{A}}\sqrt{1-t}/t = {\rm LambertW(2)}$, where, in the case of $K_{\rm S}/K_{\rm L}=1$ \cite{Zhitomirski-Dao},     
$$
{\cal{A}}\equiv\frac{\Phi_0}{4 \pi^{5/2}} \left\{\frac{\xi(0) d}{\lambda(0)}\right\} \alpha k \sqrt{
\frac{2 \gamma x}{(x + 1/x)^2} }, 
$$
and $x\equiv|\Delta_{\rm S}(0)|/|\Delta_{\rm L}(0)|$. We see from Fig. \ref{fig:t_dec} that  $t_{dec}\equiv T_{dec}/T_c$ depends on ${\cal {A}}$ strongly when ${\cal{A}} \sim {\cal{O}}(1)$. By using the parameters in Table \ref{tab:table1} which would be relevant for MgB$_2$, we obtain ${\cal{A}} \simeq 5.7$ and $t_{dec}\simeq 0.97$.

We discuss BKT transition temperature associated with the usual vortex ($\theta_{\rm L}=\theta_{\rm S}$) obtained from $k_{\rm B} T_{\rm BKT}=\Phi_0^2/32 \pi^2 \lambda_{eff} (T_{\rm BKT})$ (See, Eq. (\ref{London-limit})). By using the parameters in Table \ref{tab:table1} , we see $t_{\rm BKT} \simeq 0.90$, which is lower than $t_{dec}$ we have obtained. The present argument for $t_{dec}$ is relevant for the dilute vortex system but does not include the multi-vortex effect which would be important above $T_{\rm BKT}$, since a lot of thermally activated vortices and anti-vortices exist above $T_{\rm BKT}$ even in the system under the applied magnetic field considered here \cite{Doniach-Huberman}.  Then, the result would be corrected in this case. It should be noted that the coefficient ${\cal{A}}$ includes $\xi(0)$ explicitly, on the other hand, the equation for $T_{\rm BKT}$ does not. So, $T_{dec}$ can be lower than $T_{\rm BKT}$ in the system with sufficiently small $\xi(0)$ and the present argument can be precise. It is well known that high-T$_c$ cuprates have rather small coherence length in the order of 10$\AA$. Therefore a two-gap superconductor Ca-doped YBCO \cite{Ca-YBCO} seems to be one of the best candidates to apply the present argument. 

We comment on the relation to the renowned decoupling transition \cite{Rodriguez}. The renormalization group (RG) approach of the 2D sine-Gordon model discussed in Ref. \cite{Rodriguez} suggests that the cosine term in Eq.(\ref{sine-Gordon}) becomes irrelevant above a certain temperature $T_*$, which indicates that the decoupling between $\theta_{\rm L}$ and $\theta_{\rm S}$ occurs. The lower bound of $T_*$ is 
obtained from $T_*=8 \pi K(T_*) d$ in the limit $\gamma \rightarrow 0$ \cite{Rodriguez}. This RG analysis takes into account string loops (closed strings) only,   
i.e., it is assumed that the relative phase $\theta_{\rm L}-\theta_{\rm S}$ is single valued. 
The open string (See, Fig. \ref{fig:st-ex}) should be taken into account since it would be created by the thermal fluctuations. Actually, we have shown that 
the open string is stabilized by the entropic effect and the deconfinement occurs at $T_{dec}$.  
For sufficiently small ${\cal{A}}$, $T_{dec}$ can be lower than the lower bound value of $T_*$, i.e., the deconfinement can occur even in the temperature region that the cosine term is relevant. 

We also find that 
\begin{equation}
\frac{a(T)}{a(0)}=\left[1 - \frac{2 t}{{\cal{A}}\sqrt{1-t}} \exp\left\{-{\cal{A}}\sqrt{1-t}/t \right\}\right]^{-1},
\end{equation}
where $a(0)=2 \pi K(0)/\epsilon_{\rm st}(0)$. 
The plot of $a(T)$ is given in Fig. \ref{fig:a(T)} for ${\cal{A}}=1.6$. We see that even in the confined phase it seems to be possible to observe each of fractional vortices in the region of large $a(T)$ by using some appropriate vortex imaging techniques.

\acknowledgments
The authors are grateful to J. Akimitsu, E. Babaev, G. Baskaran, G. Blatter, H. 
Fukuyama, M. Hayashi, 
I. Herbut, T, Ishida, R. Ikeda, K. Ohashi, and Y. Tanaka for their 
critical comments. 
This work is financially supported by Grant-in-Aid for Scientific Research 
from Japan Society for the Promotion of Science under Grant 
Nos.15540370, 16740226, and 18540381.


\begin{thebibliography}{0}
\bibitem{notice} The fractional vortex is defined as a vortex 
with a flux less than $\Phi_0/2$ in Ref. \cite{Bailey-et-al}, while less 
than $\Phi_0$ in this paper.
\bibitem{note-a} Actually, $a(T)$ would have a cut off determined by the inverse of the 
square root of the applied magnetic field.
\end{thebibliography}
\end{document}